\documentclass[a4paper,10pt,pre,twocolumn,showpacs,aps,floats,superscriptaddress]{revtex4}

\usepackage{epsfig}

\begin{document}

\title{Improved routing strategies for Internet traffic delivery} 

\author{Pablo Echenique}

\affiliation{Departamento de F\'{\i}sica Te\'orica, Universidad de
Zaragoza, Zaragoza 50009, Spain.}

\affiliation{Instituto de Biocomputaci\'on y F\'{\i}sica de Sistemas
Complejos, Universidad de Zaragoza, Zaragoza 50009, Spain}

\author{Jes\'us G\'omez-Garde\~nes}

\affiliation{Departamento de F\'{\i}sica de la Materia Condensada,
Universidad de Zaragoza, Zaragoza 50009, Spain}

\affiliation{Instituto de Biocomputaci\'on y F\'{\i}sica de Sistemas
Complejos, Universidad de Zaragoza, Zaragoza 50009, Spain}

\author{Yamir Moreno}

\affiliation{Instituto de Biocomputaci\'on y F\'{\i}sica de Sistemas
Complejos, Universidad de Zaragoza, Zaragoza 50009, Spain}

\affiliation{Departamento de F\'{\i}sica Te\'orica, Universidad de
Zaragoza, Zaragoza 50009, Spain.}

\date{\today}

\widetext

\begin{abstract} 

We analyze different strategies aimed at optimizing routing policies
in the Internet. We first show that for a simple deterministic
algorithm the local properties of the network deeply influence the
time needed for packet delivery between two arbitrarily chosen
nodes. We next rely on a real Internet map at the autonomous system
level and introduce a score function that allows us to examine
different routing protocols and their efficiency in traffic handling
and packet delivery. Our results suggest that actual mechanisms are
not the most efficient and that they can be integrated in a more
general, though not too complex, scheme.

\end{abstract}


\maketitle

Modern society increasingly depends on large communication networks
such as the Internet. The need for information spreading pervades our
lives and its efficient handling and delivery is becoming one of the
most important practical problems. To this end, a suite of protocols
for the dissemination of information from a given source to thousands
of users has been developed in the last several years
\cite{alexvb}. The Internet and other communication networks certainly
work in a reliable way. However, both the physical network and the
numbers of users are growing continuously. The scalability of current
protocols as well as their performance for larger system sizes and
heavier loads on the network are critical issues to be addressed
in order to guarantee networks' functioning in the near future. From
this perspective, one of the fundamental problems we face nowadays is
to find optimal strategies for packet delivery between a given sending
node and its destination host.

The above problem demands taking into account at least two
factors. The first is related to the routers and servers capabilities
$-$ mainly, memory requirements and CPU processing time $-$ needed for
the different algorithms to operate efficiently. Secondly, it has been
recently shown that the real architecture of communication networks
determines many of their properties in front of dynamical processes
such as the resilience to random failures and attacks and the
spreading of virus and rumors
\cite{havlin01,newman00,moreno02,av03,n02b}. The latter has been
achieved in recent years by unraveling the complex patterns of
interconnections that characterize seemingly diverse systems such as
the Internet, the World Wide Web (WWW), biological and social networks
\cite{strogatz,book1,book2}. It turns out that most real networks can
be described by growing models in which the number of nodes (or
elements) forming the net increases with time and that the probability
that a given node has $k$ connections to other nodes follows a power
law $P_k\sim k^{-\gamma}$. This is the case of the Internet which
shows a scale-free (SF) connectivity distribution with an exponent
that has been estimated to be around $\gamma=2.2$ \cite{alexvb}.

The aim of this paper is twofold. First, we show that the local
properties of the networks on top of which a packet delivery process
is taking place determine its efficiency. Then, we turn our attention
on developing alternative strategies for Internet traffic routing by
implementing a general stochastic protocol on top of a real Internet
map at the autonomous system (AS) level. Through Monte Carlo
simulations we show that the actual Internet's protocol is less
efficient than one which combines the knowledge of the topology and
the traffic handled by the network at a local scale \cite{nn1}. Our
results might provide new hints for routing protocol designs as they
are shown to need roughly the same capabilities of current routers.

In order to discuss how the local topological properties influence the
efficiency of a given routing protocol, we use the network studied in
Ref.\cite{jesus}. In this model, the network is generated by
considering the Barab\'asi-Albert (BA) procedure \cite{bar99} but
introducing an affinity variable $f_i$ and a tolerance $\mu$, which
determine the peers $j$ a new node can attach to. This is done by
requiring that $f_j\; \in\; (f_i\pm\mu)$. This network shows the same
global properties of the BA net regardless of the tolerance. However,
depending on the value of $\mu$, other local properties, such as the
clustering coefficient and correlations, differ from the original BA
network.

Once we have a network whose local properties can be tuned, we proceed
to define our basic routing protocol. In communication networks like
the Internet, routers deliver data packets by ensuring that all
routers converge to a best estimate of the path leading to each
destination address. In other words, the routing process takes place
following the criterion of the shortest available path length from a
given source to its destination. Inspired by this protocol and with
the aim of comparing its performance with other strategies, we define
the following algorithm for packet delivery, which will be called the
{\em standard} protocol. At the beginning, $p$ packets are created and
both their destinations and the sources are chosen at random. In
subsequent time steps, each node $i$ holding a packet sends it to its
destination $j$ following the shortest path length between node $i$
and $j$ until all packets reach their destinations. That is, each
packet is diverted in such a way that the distance $d_{ij}$, measured
as the number of nodes one needs to pass by between $i$ and $j$, is
minimized. In the case that there are more than one possible path, the
choice is made at random. This recipe has been recently studied in
generic network models \cite{t1,t2}

\begin{figure}[t]
\begin{center}
\epsfig{file=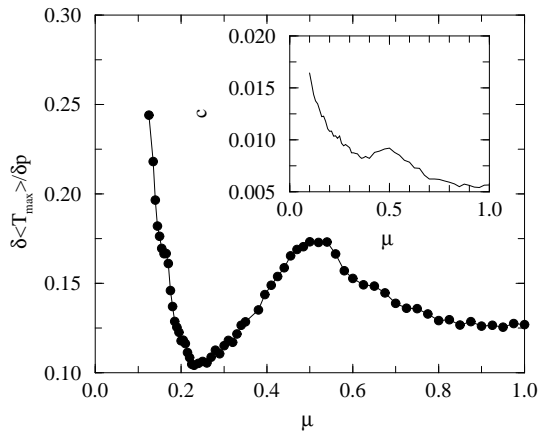,width=2.3in,angle=-90,clip=1}
\end{center}
\caption{$\partial \langle T_{max}\rangle /\partial p$, as a function
of the network parameter $\mu$. The figure illustrates the dependency
of the standard routing protocol on the local properties of the
network. The inset shows the variation of $c$ with $\mu$. The size of
the network is $N=10^4$ nodes and $m_o=m=3$. The degree distribution
is a power law with exponent equal to $3$. Note that the BA limit
corresponds to $\mu=1$. See the text for further details.}
\label{figure1}
\end{figure}

The above procedure is repeated many times for a number of processes
ranging from $p=1$ to at least $p=500$. Different realizations of the
dynamics for the same $p$ are performed in order to average the
relevant quantities. As a measure of the {\em efficiency} of the
process, we have monitored $\langle T_{max} \rangle$, the maximum time
it takes for a packet to travel from its source to its destination,
averaged over different realizations \cite{nn2}. The numerical results
show that this magnitude scales linearly with the number of processes,
so that its derivative is a proper order parameter to characterize the
routing performance. Figure\ \ref{figure1} shows the slopes of the
straight lines as a function of the control parameter $\mu$ which
determines the local properties of the network. It is clear from the
figure that the algorithm's outcome depends on the topological details
of the network. For this network the average shortest path length $L$
is roughly the same as that of the BA network up to a value of $\mu$
around $0.2$. However, in this range of $\mu$ the efficiency has a
maximum and a minimum. This implies that other properties are
responsible for the behavior observed, namely the clustering
coefficient $c$. For the network under analysis \cite{jesus}, it turns
out that as $\mu$ is decreased from 1 (the BA limit) to smaller
values, $c$ first slightly decreases below the BA value (accounting
for the maximum) and then starts increasing up to values near the ones
of real networks (see the inset in Fig.\ \ref{figure1}). From this
perspective, it is then clear that the routing process is controlled
by $c$ in the region where $L$ is constant. When $c$ increases, there
appear new connections among the neighbors of a node and the number of
shortest paths raises. Now, packets can circumvent more easily
cogested nodes, thus making the standard protocol more efficient. When
$c$ decreases, the opposite occurs. For very small $\mu$, $L$ diverges
\cite{jesus} leading to a bad performance of the protocol since the
algorithm works on a shortest-path-delivery basis. The crossover from
the minimum to the divergence of $\partial \langle T_{max}\rangle
/\partial p$ is achieved in the parameter region where the interplay
between $c$ and $L$ breaks down.

The preceding analysis shows that the routing protocol may be very
sensitive to local details of the network on top of which the
spreading process is taking place. It is then advisable the use of
real nets in order to obtain reliable results. To this end, we have
used the Internet Autonomous System map at the Oregon route server
dated May 25, 2001 \cite{as}. It is worth stressing that each AS
groups many routers together and the traffic carried by a node is the
aggregation of the traffic generated at the internal routers and on
individual end-host flows between the ASs.

The first modification of the routing mechanism is introduced by
noting that the standard procedure does not take into account the
traffic on the network. Specifically, a routing policy based on the
shortest path between two given nodes neglects the queue in overloaded
nodes which makes the process slower as the queue lengths become
larger. That is, it may be more efficient to divert a packet through a
larger but less congested path. Let us hence assume that a node $l$ is
holding a packet that should be sent to a node $j$ and define an
effective distance $d_{eff}^i$ from a neighboring node $i$ of $l$ to
the destination $j$ as $d_{eff}^i=d_i+c_i$, where $d_i$ is the
shortest path between node $i$ and $j$ and $c_i$ is the number of
processes (or packets) in the queue of $i$. The above definition,
however, does not allow us a direct comparison with the standard
procedure. It is convenient to redefine the effective distance as
$d_{eff}^i=h_d d_i+ h_c c_i$ so that the limit $h_c=0$ contains the
standard protocol. Furthermore, without loss of generality, we assume
$h_d+h_c=1$. This algorithm will be called {\em deterministic}
protocol henceforth. The rest of the algorithm remains the same as
before, i.e., at each time step, the remaining packets are delivered
in such a way that the path chosen is that which minimizes
$d_{eff}^i$. 

A first look at the dynamics shows that a protocol implemented in this
way is more efficient than taking into account only the shortest path
criterion. In fact, $\langle T_{max} \rangle$ departs from the linear
behavior previously observed and is well below the straight line up to
a high $p$. This behavior clearly depends on $h_d$, since it is
straightforward to realize that if $h_d$ is zero, the packets are
diverted following the less loaded node regardless of the path length
which results in an uncontrolled increase in the distance traveled by
the packets from the sending nodes.

\begin{figure}[t]
\begin{center}
\epsfig{file=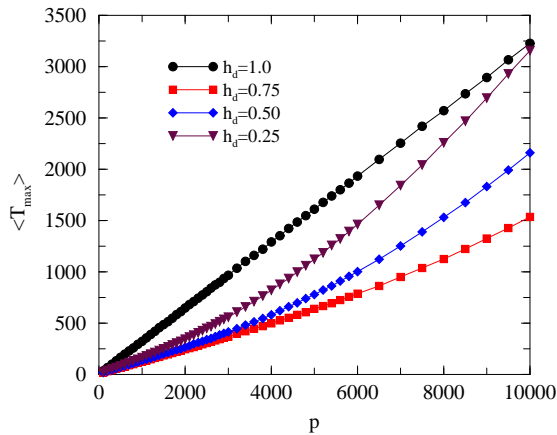,width=2.3in,angle=-90,clip=1}
\end{center}
\caption{(Color online) Dependency of $\langle T_{max} \rangle$ on the
  number of initial packets $p$ in the deterministic limit of the
  model ($\beta=20$) ran on top of an AS Internet map made up of
  around 11000 nodes. Each point is an average over at least 200
  realizations. The standard protocol corresponds to the limit
  $h_d=1$. Note that although the tendency of the curves is to cross
  the straight line as $p$ increases, there is an optimal value of
  $h_d$ such that the interception would take place in the limit of
  very heavy traffic.}
\label{figure2}
\end{figure}

The above algorithm can be further generalized by including a
probabilistic view. In other words, once we have determined the
$d_{eff}^i$ for all sending nodes $i$, we can allow for a stochastic
choice of the paths. Hence, our third algorithm, referred to as {\em
stochastic} protocol considers a score function or ``energy'' $H_i=h_d
d_i+ (1-h_d) c_i$ and that the probability $\Pi_i$ that a packet is sent
precisely through node $i$ is given by,
\[
\Pi_i=\frac{e^{-\beta H_i}}{\sum e^{-\beta H_j}}
\]
where $\beta$ is the inverse of the temperature. In the limit
$\beta\rightarrow\infty$ (at zero temperature) we recover the
deterministic protocol.

Figure\ \ref{figure2} shows the dependency of $\langle T_{max}
\rangle$ on the number of packets $p$ for several values of $h_d$ in
the deterministic limit of the model, which we found to be fulfilled
for $\beta=20$. A dynamics which does not take into account the amount
of traffic handled by the neighbors of a sender node $-$straight line
in Fig.\ \ref{figure2}$-$ performs worse than the one which integrates
both ingredients. However, this depends on the specific weight of each
metric in $H_i$ and on $p$. In the regime where the traffic is not
heavy (small $p$ values) all curves are below the standard protocol
performance, but as the amount of traffic handled by the network
increases, the deterministic protocol starts performing worse for a
range of $h_d$ values. From the results, it seems that eventually,
when the traffic increases too much, the curves cross the straight
line indicating that at those limits the shortest path strategy is
best suited. Note, however, that for $h_d=0.75$ the convergence of the
two algorithms occurs for a very heavy load. Consequently, we can
assert that there is an $h_d$ region where the combination of the two
ingredients gives rise to the best performance. On the other hand, the
existence of an optimal $h_d$ value distinct from zero can be
understood by noting that a mechanism lacking some degree of path
length information between the source and destination nodes of the
packets performs badly because the packets travel along too large
paths that do not compensate the time they would loose trapped in the
queues of congested nodes.

\begin{figure}[t]
\begin{center}
\epsfig{file=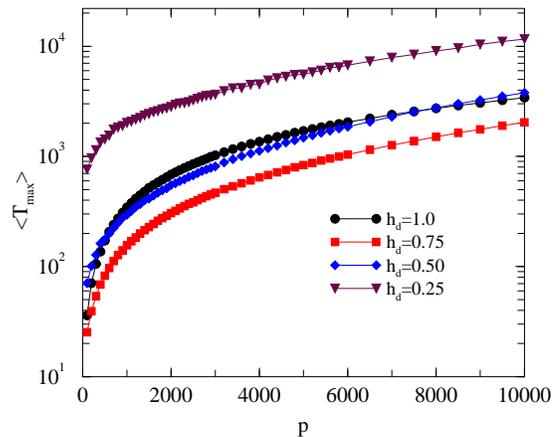,width=2.3in,angle=-90,clip=1}
\end{center}
\caption{(Color online) Dependency of $\langle T_{max} \rangle$ on the
  number of packets, $p$, for a middle $\beta=5$ value. The network
  parameters are as in Fig.\ \ref{figure2}. In this case, the $h_d$
  range in which the stochastic strategy performs better than the
  standard one is reduced.}
\label{figure3}
\end{figure}

The completely stochastic limit of the model corresponds to
$\beta=0$. The performance of the protocol in this limit is however
very bad. In fact, for an infinite temperature, all neighboring nodes
of a given sender have the same probability to receive the message,
and the dynamics becomes a random walk process. With no topological
information about what are the destinations of the packets, they
arrive to the receiver at longer times and the algorithm is the
worst. For intermediate values of $\beta$, we have an stochastic
dynamics in which topological and traffic information coexist. This is
the case depicted in Fig.\ \ref{figure3} for the same values of $h_d$
used in Fig.\ \ref{figure2}. As can be noted from the figure, the
stochastic protocol increases $\langle T_{max} \rangle$ by at least
one order of magnitude as compared to the deterministic limit
($\beta=\infty$). Moreover, the standard approach seems to be the best
choice for a wider range of $h_d$ values, although $h_d=0.75$ still
performs better.

\begin{figure}[t]
\begin{center}
\epsfig{file=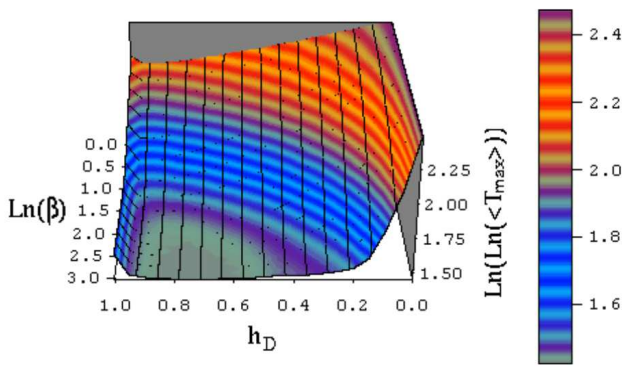,width=3.3in,angle=0,clip=1}
\end{center}
\caption{(Color online) Phase diagram of the system's dynamics. The
network parameters are as in Fig.\ \ref{figure2}. The number of
processes is $p=500$. Calculations for higher $p$ show that the
minimum of $\langle T_{max} \rangle$ is also attained around
$h_d=0.8\pm 0.1$.}
\label{figure4}
\end{figure}

Figure\ \ref{figure4} summarizes our results for different values of
the control parameters $\beta$ and $h_d$. It turns out from the study
of the whole phase diagram that the best algorithm is one which
includes information about both path lengths and congestion at a local
scale. Besides, the deterministic limit with $h_d=0.75$ gives the best
results for $\langle T_{max} \rangle$. It would be worth noticing at
this point that, although the Fig.\ \ref{figure4} was obtained in not
too heavy traffic conditions, the results are consistent for larger
values of $p$. Different tests allow us to conclude that the optimal
value is $h_d=0.8\pm 0.1$. In any case, this confirms that it would be
possible to device more elaborated protocols with the aim of
diminishing the time needed for a packet to spread through the
network. In light of the present results, such an strategy may be
implemented by also taking into account the amount of traffic handled
by a local area of the network.

As suggested by Fig.\ \ref{figure4}, the best protocol is the
deterministic one, which, on the other hand, should be more easy to
implement in practice. The microscopic dynamics of the routing process
in this limit reveals that it is desirable that the routing process
incorporates some knowledge of the node's queue lengths. However, the
contribution in the score function of the second term should not weigh
in excess. For small values of $h_d$, say 0.25, the algorithm performs
better that the standard one because the packets do not pass by the
hubs of the network, which are likely to be in the shortest path route
to any node. Instead, they go around the hubs and $\langle T_{max}
\rangle$ is smaller. If $p$ is increased, the neighbors of the hubs
also get congested. This leads to a situation in which the packets
around a hub get trapped in its neighborhood, getting in and out from
it, but without finding their routes to their destinations.

We finally point out that the improved strategies studied in this
paper should not be hard to implement in practice. Actual protocols
use the topological information used by the different variants
explored. One would only need to provide further information on the
traffic status in a local area. This could be done, for instance, by
using the {\em keep-alive} messages that routers continuously exchange
with their peers, though in this case there is some time delay due to
the time scales of the routing process and the database or message
exchanges.

In summary, we have studied alternative strategies for traffic
delivery in complex heterogeneous networks. The results showed that
the performance of the standard approach is sensitive to local
topological changes. Specifically, the clustering properties may play
a key role in message delivery. On the other hand, algorithms which
integrate topological and traffic information have been shown to
perform better that the standard protocol. Finally, we note that the
present results could be extended to the Internet map at the router
level which is statistically equivalent \cite{alexvb} to the AS map
used here.


\begin{acknowledgments}
Y.\ M.\ thanks A. V\'azquez for sharing his Internet database with
us. P.\ E.\ and J.\ G-G\ acknowledge financial support of the MECyD
through FPU grants. Y.\ M.\ is supported by a BIFI Research
Grant. This work has been partially supported by the Spanish DGICYT
project BFM2002-01798.
\end{acknowledgments}


\begin{thebibliography}{99}
\bibitem{nn1} Several traffic-dependent routing protocols with rules
  different to those employed here have been studied by computer
  scientists without taking into account the actual topology of
  real-world nets.
\bibitem{nn2} The choice of $\langle T_{max} \rangle$ is
  arbitrary. One can also use $\langle T_{avg} \rangle$ or $\langle
  T_{rms} \rangle$, which lead to the same behaviors. We have chosen
  $\langle T_{max} \rangle$ because the errors involved in its
  computation are the smallest of the three regardless of the
  algorithm.
\end{thebibliography}
\end{document}